\documentclass[preprint,showpacs,showkeys,preprintnumbers,amsmath,amssymb]{revtex4}

\usepackage{natbib}
\usepackage{graphicx}% Include figure files
\usepackage{dcolumn}% Align table columns on decimal point
\usepackage{bm}% bold math
\usepackage{longtable}% long table

\begin{document}

%\preprint{APS/123-QED}

\title{Does Spin-Orbit Coupling Effect Favor Planar Structures for Small Platinum Clusters?}

\author{Ali Sebetci}
 \email{a.sebetci@ifw-dresden.de}
\affiliation{IFW Dresden, Helmholtzstrasse 20, 01069 Dresden,
Germany}

\date{\today}% It is always \today, today,

\begin{abstract}

We have performed full-relativistic density functional theory 
calculations to study the geometry and binding energy of 
different isomers of free platinum clusters Pt$_{n}$ ($n=4-6$) 
within the spin multiplicities from singlet to nonet.  
The spin-orbit coupling effect has been discussed for the 
minimum-energy structures, relative stabilities, 
vibrational frequencies, magnetic moments, and the highest 
occupied and lowest unoccupied
molecular-orbital gaps. It is found in contrast 
to some of the previous calculations that 3-dimentional 
configurations are still lowest energy structures of these
clusters, although spin-orbit effect makes some planar 
or quasi-planar geometries more stable than some other
3-dimentional isomers. 
												 
\end{abstract}

%\keywords{}

%Use showkeys class option if keyword
\maketitle

\section{INTRODUCTION}

Small transition metal clusters have been attracting wide 
interest due to their potential applications as building 
blocks of functional nanostructured materials, electronic 
devices, and nanocatalysts~\cite{Schmid}.
In particular, platinum is one of the most important ingredients 
in the heterogeneous catalysis of hydrogenation as well
as in the catalysis of the CO, NO$_{x}$, and hydrocarbons. 
It is currently the preferred oxidation and
reduction catalyst for low-temperature PEM fuel 
cells~\cite{Viel}. As the size of the Pt particles decrease, 
their catalytical activities tend to increase because 
of the increased surface areas of smaller
particles and the structural sensitivity of some reactions~\cite{Schneider}. 

Lineberger and co-workers have investigated the electronic 
spectra of small platinum and palladium ($n=2,3$) clusters 
by the negative ion photodetachment spectroscopic 
method~\cite{Ho,Ervin} and Eberhardt and co-workers have obtained 
the valence and core-level photoemission spectra of mass 
selected monodispersed Pt$_{n}$ ($n$=1-6) clusters~\cite{Eberhardt}. 
In the recent theoretical calculations, we have studied bare 
and hydrogenated Pt$_{n}$H$_{m}$ ($n$=1-5 $m$=0-2) clusters
in a scalar-relativistic density functional theory (DFT) 
formalism~\cite{Ali}. A relevant literature can be found in 
the reference~\cite{Ali} for the previous experimental and 
theoretical investigations. In addition to them, Saenz {\it et al}~\cite{Saenz}
have worked on the interaction of Pt clusters with molecular oxygen.
Futschek {\it et al}~\cite{Futschek} have presented ab initio density functional 
studies of structural and magnetic isomers of Ni$_n$ and Pt$_n$ clusters with 
up to 13 atoms. Seivane and Ferrer~\cite{Seivane} have analyzed the impact of 
the magnetic anisotropy on the geometric structure and magnetic ordering of
small atomic clusters of palladium, iridium, platinum, and gold from two to 
five, six, or seven atoms, depending on the element.
Bhattacharyya and
Majumder~\cite{Bhat} have reported the growth pattern and 
bonding trends in Pt$_{n}$ ($n$=2-13) clusters and concluded in their first
principles study that small Pt$_{n}$ clusters have planar geometries 
and a structural transition to non-planar geometries occur at $n$=10. 
Similarly, Huda {\it et al}~\cite{Huda} predicted that spin-orbit (SO)
coupling effect favors planar structures for small Pt clusters.

In the present work, we discuss the effect of SO 
coupling on the structural, electronic and magnetic 
properties of small Pt$_{n}$ ($n$=4-6) clusters by employing 
Gaussian atomic-orbital methods in the full-relativistic
DFT. The possible local minima and ground state isomers, 
binding energies (BE), relative stabilities, magnetic moments, 
and the highest occupied and the lowest unoccupied 
molecular-orbital (HOMO-LUMO) energy gaps have been
calculated with and without SO coupling to compare the 
results for both cases. Vibrational frequency calculations 
for each optimized structure has been carried out to 
differentiate local minima from transition states. 

\section{COMPUTATIONAL METHOD}

NWChem 5.0 program package~\cite{NWChem} has been used to 
perform geometry optimizations and total energy
calculations by DFT. CRENBL~\cite{Ross} basis set, effective 
core potential and SO operator for Pt have been employed 
where the outer most 18 electrons of free Pt atom 
(5s$^{2}$5p$^{6}$5d$^{9}$6s$^{1}$) are treated as valence 
electrons. The generalized gradient approximation of Becke's exchange
functional~\cite{Becke} and Lee-Yang-Parr correlation 
functional~\cite{Lee} (B3LYP) has been chosen as the 
hybrid exchange-correlation (xc) functional. Wave function 
and geometry optimization convergence criteria were not worse than the default
NWChem criteria. The geometries have been relaxed without 
imposing any symmetry constraints. Spin-polarized calculations 
have been done for the first five spin multiplicities 
(from singlet to nonet). Figures were produced by ChemCraft graphics 
program~\cite{ChemCraft}.

\section{RESULTS AND DISCUSSION}

\subsection{Pt and Pt$_{2}$}

First, we discuss the properties of Pt and Pt$_{2}$ to 
assess the accuracy of our theoretical method.
The ground-state Pt atom was found in the triplet state 
(5d$^{9}$6s$^{1}$) for the non-SO coupling case in agreement 
with the experimantal results~\cite{Moore}. The excitation 
energy of the singlet 5d$^{10}$6s$^{0}$ state has been calculated as
0.510 eV (neglecting SO coupling) which can be compared 
with a spin-averaged experimental value of 0.478 eV~\cite{Moore}. 
By including SO effects, the excitation energy of the 
closed-shell configuration is 0.881 eV to be compared with the experimental 
value of 0.761 eV~\cite{Moore}. Both of these calculated 
excitation energies are much better than the ones obtained by Fortunelli 
in 1999~\cite{Fortunelli}. We have calculated the SO splitting 
in the 5d orbital of Pt atom in its singlet state (5d$^{10}$6s$^{0}$) 
for simplicity and obtained as 1.131 eV, which is in a 
reasonable agreement with the value of 1.256 eV calculated 
by the all-electron full-relativistic DFT code FPLO~\cite{FPLO} 
by employing Perdew-Wang~\cite{Per-Wang} local xc 
functional in both programs. 
The ionization potential (IP) of Pt has been calculated as 9.319 eV 
in the full-relativistic case where the experimental 
value is 8.958 eV~\cite{Marij}. Huda {\it et al} has calculated 
IP of Pt as 9.381 eV in their DFT study with the 
projected augmented wave (PAW )method~\cite{Huda}. The calculated 
electron affinity is 1.806 eV which is comparable to the experimental 
value of 2.123$\pm$0.001 eV~\cite{Gibson}. The discrepancy 
between the calculated and experimental values of the electron 
affinity may be reduced when the experimental value is improved 
by SO effects which can be 
estimated to decrease it by about 0.2 eV~\cite{Hotop}. 

The bond lengths for the Pt dimer were found to be 2.373 $\AA$ 
and 2.406 $\AA$ for non-SO and SO calculations respectively,  
while the experimental value~\cite{Airola} is 2.333 $\AA$. 
The corresponding binding energies are 2.708 eV (non-SO) and
2.476 eV (SO), while the experimental value is 3.14$\pm$0.02 
eV~\cite{Taylor}. Thus, the present method slightly overestimates
the Pt dimer bond length and therefore underestimates the 
binding energy. Both of the calculated vibrational frequencies 
of 237 cm$^{-1}$ (non-SO) and 219 cm$^{-1}$ (SO) agree pretty 
well with the measured value of 223 cm$^{-1}$~\cite{Fabbi}.  

\subsection{Pt$_{4}$}

\begin{figure}
\includegraphics[bb= 0 0 8.5cm 13.2cm, scale=1]{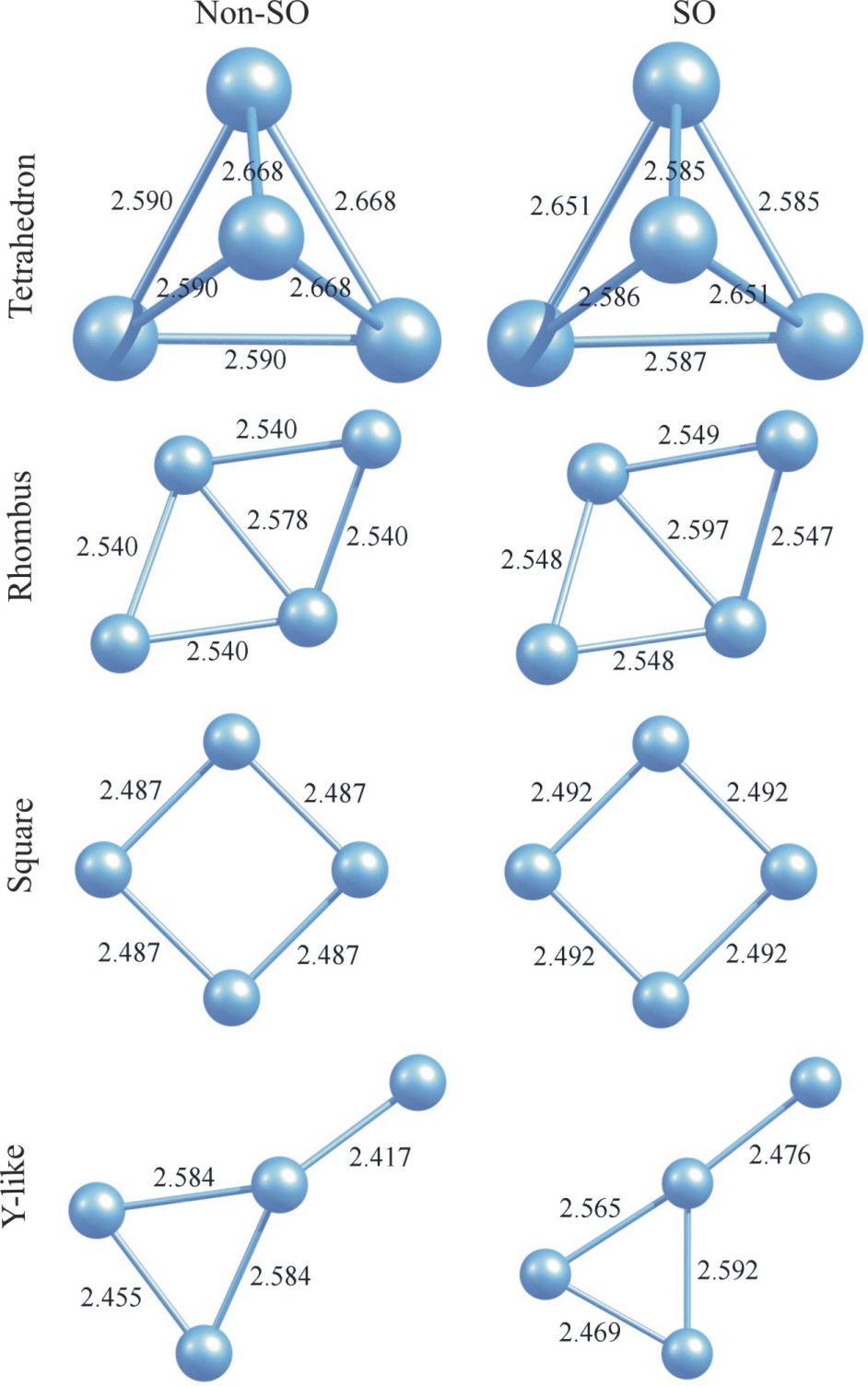}
\caption{\label{pt4}Relaxed geometries of Pt$_4$ isomers for the most stable spin multiplicity of each isomer with
and without SO coupling effects (distances are in $\AA$).}
\end{figure}

We report the relative stability of Pt$_{4}$ isomers with and without 
SO coupling for different spin multiplicities in 
Table~\ref{table-pt4}. The geometric structures and bond lengths of each isomer of Pt$_{4}$ 
for their most stable spin multiplicities 
are given in Fig.~\ref{pt4}.  

Similar to the results in our previous study on bare and hydrogenated 
small Pt clusters~\cite{Ali} where we did not employ SO coupling
effects and to the references~\cite{Bala,Futschek}, the calculated lowest 
energy structure of Pt tetramer is a tetrahedron for both cases of non-SO and SO. 
When SO effects are not 
considered, the ground state spin multiplicity of the tetrahedron 
is 3 and it has the point group symmetry $C_{3v}$. 
On the other hand, when SO effects are taken into account, quintet, 
septet, and nonet initial spin multiplicities converge to the same spin moment of 3.64 $\mu_{B}$ and 
this state has 26 meV lower energy than the one with 1.84 $\mu_{B}$ spin moment. 
In addition, the point 
group symmetry of the optimized structure in this case
is $D_{2d}$. Our prediction of tetrahedron as the ground state 
geometry is in contrast with some of the previous 
results~\cite{Bhat,Huda,Gronbeck}. A common feature of these studies 
is that they have been done by employing plane wave codes
(references \cite{Bhat,Huda} are by PAW method of VASP~\cite{Vasp}, 
reference \cite{Gronbeck} is by CPMD~\cite{Cpmd}) and predict that the 
rhombus isomer is the lowest energy structure with a spin multiplicity 
of 5 despite the fact that Futschek {\it et al}~\cite{Futschek} have employed 
the code VASP too and obtained a distorted tetrahedron similar to our results. 
According to the present calculations of non-SO case, the quintet state 
is the ground magnetic state of the rhombus, but its total energy 
is 0.206 eV higher than that of the tetrahedron. For the SO case,
the rhombus whose singlet, triplet, quintet, and septet initial spin multiplicities converge to
a spin moment between 2.08 and 2.11 $\mu_{B}$ has 0.133 eV higher energy than the tetrahedron. 
The optimized
structures of the rhombus in both non-SO and SO cases have the same 
point group symmetry $C_{2v}$. As in the case of the dimer, SO effects 
slightly stretch the bond lengths. Both of these structures are out 
off plane where the angles between the triangular planes are 113$^{\circ}$ 
(non-SO) and 105$^{\circ}$ (SO). Thus, SO effect strengthens the non-planarity of 
the rhombus in contrast to the findings of Huda {\it et al}~\cite{Huda}. The third,
and fourth lowest energy isomers of Pt tetramer are the square and Y-like 
(see Fig.~\ref{pt4}) planar structures, respectively. The total energy
of the singlet square (non-SO) is 0.232 eV higher than that of the 
global minimum. SO effect increases this energy difference to a value of 
0.357 eV which is dissimilar to the general trend that SO effects 
decrease the energy differences between the isomers. 
For the Y-like isomer, the energy seperations from the 
lowest energy structure 
are 0.551 eV (non-SO quintet) and 0.337 eV (SO 3.47 $\mu_{B}$).  

\subsection{Pt$_{5}$}

\begin{figure}
\includegraphics[bb= 0 0 10.0cm 20.7cm, scale=1]{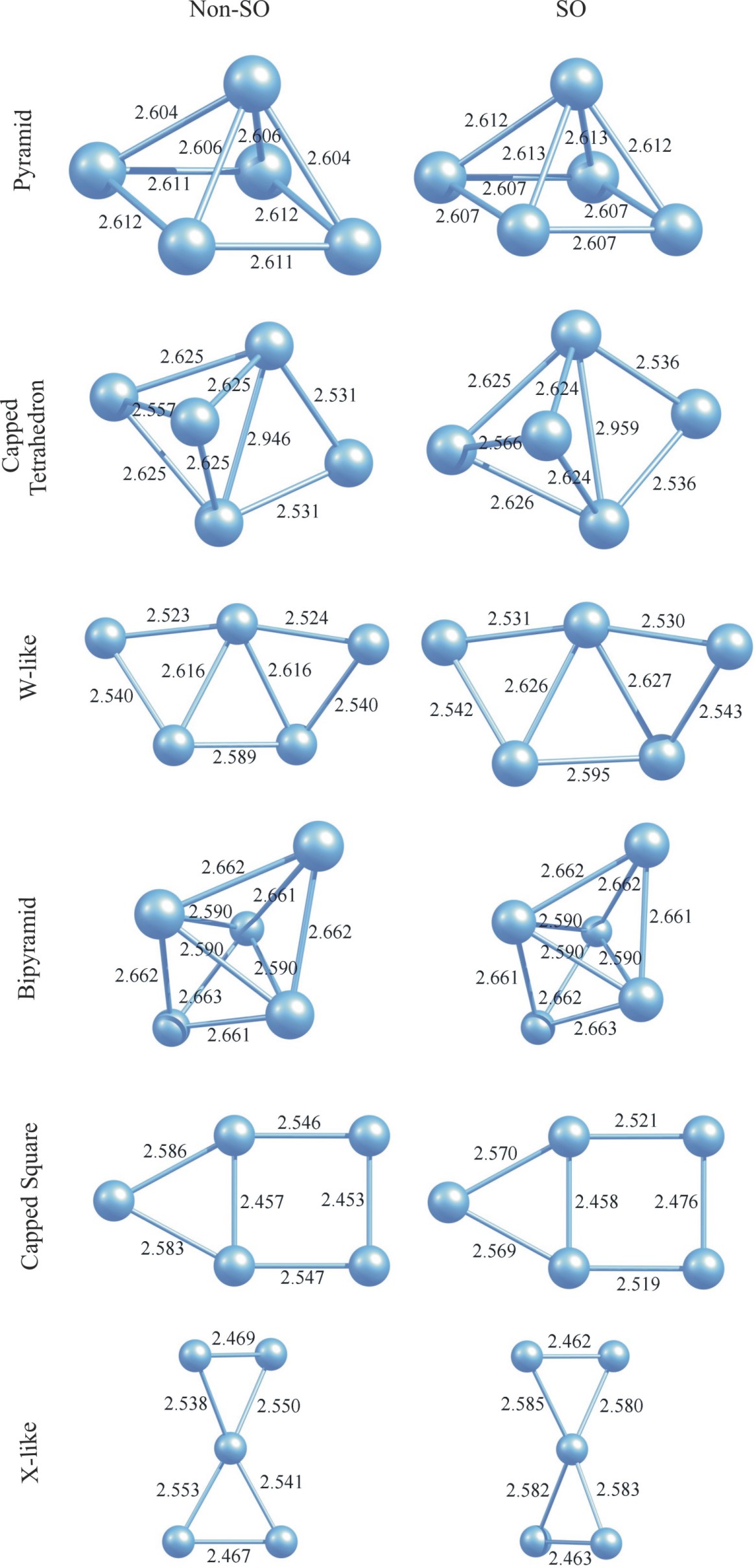}
\caption{\label{pt5}Relaxed geometries of Pt$_5$ isomers for the most stable spin multiplicity of each isomer with
and without SO coupling effects (distances are in $\AA$).}
\end{figure}

We have identified six different stable isomers of Pt$_{5}$ clusters 
and reported their relative stabilities in Table~\ref{table-pt5}
and their structures and bond lengths in Fig.~\ref{pt5}. Although 
a bridge side capped tetrahedron with a spin multiplicity of 5 
has been obtained as the lowest 
energy structure for the non-SO case which agrees with our previous 
calculations~\cite{Ali}, SO coupling effects favor a pyramid with 5.58 $\mu_{B}$
spin moment. 
The energy seperation between the capped tetrahedron and the pyramid 
in the former case is only 11 meV which
may be considered within the accuracy of the calculations. On the other 
hand, SO effects favor pyramidal structure against the 
tetrahedral one as much as 167 meV. While the obtained pyramidal isomer 
as the lowest energy structure contradics the findings of 
Bhattacharyya and Majumder~\cite{Bhat}, Huda {\it et al}~\cite{Huda}, 
Gr{\"o}nbeck and Andreoni~\cite{Gronbeck}, Xiao and Wang~\cite{Xiao},
Saenz {\it et al}~\cite{Saenz}, and Futschek {\it et al}~\cite{Futschek},
it agrees with the predictions of Balasubramanian {\it et al}~\cite{Bala}, 
Yang {\it et al}~\cite{Yang}, and Seivane and Ferrer~\cite{Seivane}. 
Bhattacharyya and Majumder~\cite{Bhat}, and
Huda {\it et al}~\cite{Huda} have predicted planar bridge side capped structure,
Gr{\"o}nbeck and Andreoni~\cite{Gronbeck}, and Saenz {\it et al}~\cite{Saenz}
have predicted planar W-like structure, Xiao and Wang~\cite{Xiao}, and
Futschek {\it et al}~\cite{Futschek} have predicted 3-dimensional trigonal
bipyramid as the global minimum configuration. 

The third low lying isomer is an out of plane W-like structure (see Fig.~\ref{pt5}) 
with a spin moment of 3.58 $\mu_{B}$ having 173 meV higher energy than the global minimum. 
Previously we have identified this trapezoidal-type structure as the fifth low
lying isomer~\cite{Ali}. However, in that study we have considered only 
its singlet and triplet states.
Thus, the discrepancy can be attributed to the limitation of the previous 
calculations for the first two spin multiplicities. 
The fourth energetically favorable structure is a bipyramid with 3.57 $\mu_{B}$ 
spin moment. The planar, bridge side 
capped square structure which has been predicted as the global minimum
of Pt$_{5}$ clusters~\cite{Bhat,Huda} by a plane wave code, has been obtained 
as the fifth isomer in our calculations. It has 0.674 eV (non-SO) and 0.373 eV
(SO) higher energies than the lowest energy structures. Finally, a non-planar 
X-like gometry has been identified as a stable isomer with the highest 
total energy. Unlike most of the other isomers, the most stable magnetic 
state of the X-like structure for SO case could be obtained from an initial
nonet state which gives very high relative energy (2.859 eV) in non-SO calculations. 

The effect of SO coupling on the bond lengths of Pt$_{5}$ clusters is not 
monotonic. For the pyramidal structure, while the interatomic distances
between the apex atom and the atoms on the squared plane are stretched, 
the distances between the atoms on the squared plane are diminished. 
For the capped tetrahedron, as SO effects make the triangle constructed 
by the capping atom and the two atoms bonded to the capping atom a bit larger,
all other bond lengths are kept nearly the same. All bond lengths of W-like 
structure and most of them in X-like structure are stretched by SO coupling, 
on the contrary, it does not have a significant effect on the bond length of 
the bipyramid, and even causes contractions in the bond length 
of the capped square isomer.   

\subsection{Pt$_{6}$}

\begin{figure}
\includegraphics[bb= 0 0 17.169cm 20.513cm, scale=0.9]{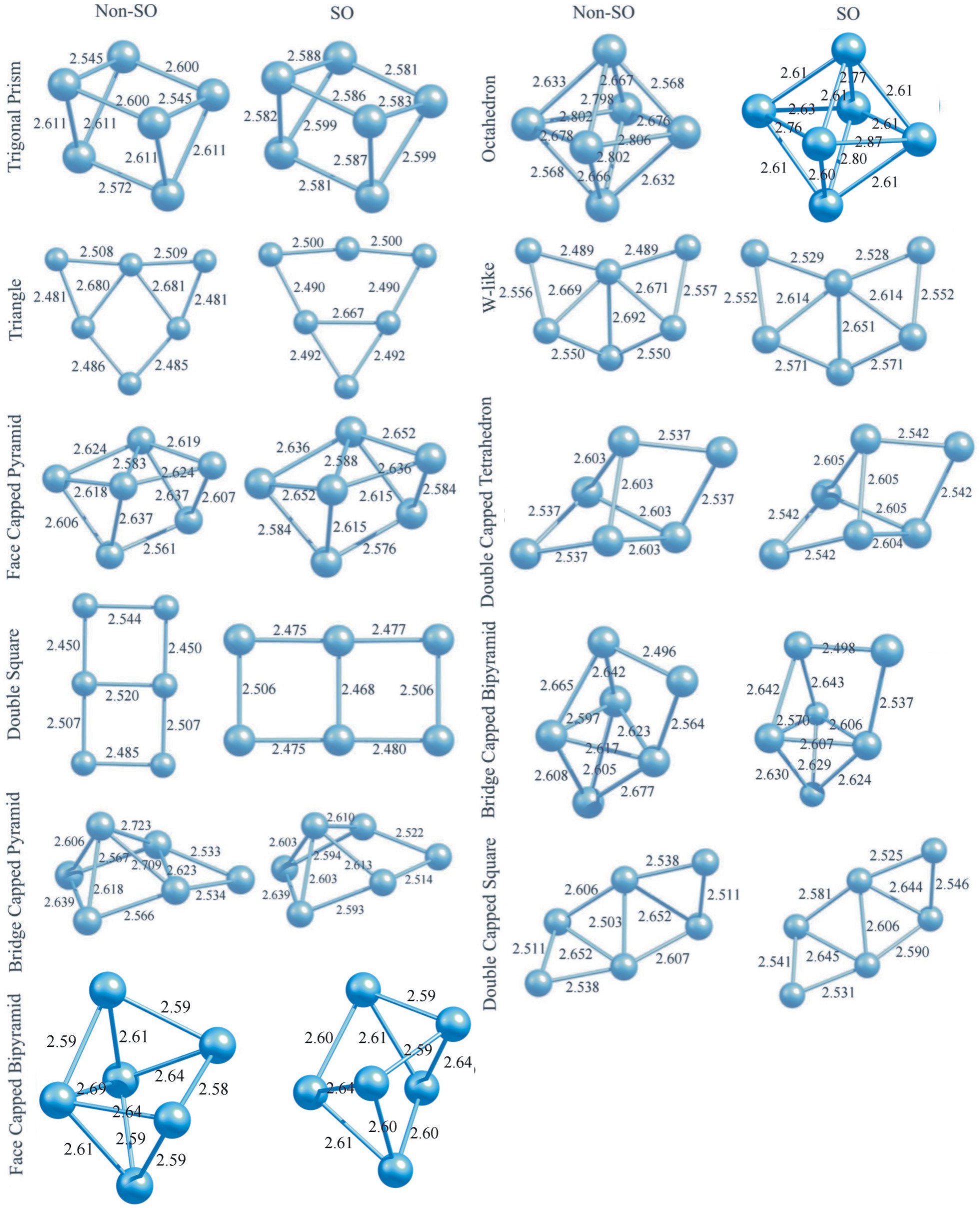}
\caption{\label{pt6}Relaxed geometries of Pt$_6$ isomers for the most stable spin multiplicity of each isomer with
and without SO coupling effects (distances are in $\AA$).}
\end{figure}

The relative stabilities of eleven different isomer of Pt$_{6}$ can be found in 
Table~\ref{table-pt6}. Their bond lengths and structures have been given in 
Fig.~\ref{pt6}. The most stable structure of Pt$_{6}$ has been obtained as a trigonal prism 
in the septet state for non-SO case and a state with 5.43 $\mu_{B}$ spin moment for SO case. 
Xiao and Wang~\cite{Xiao} 
have predicted a planar 
double capped square, Bhattacharyya and Majumder~\cite{Bhat} have identifed
a planar triangular structure, Futschek {\it et al}~\cite{Futschek} have obtained
a face capped pyramid as the lowest energy isomer of Pt$_{6}$. In our 
calculations, the triangle is the second lowest energy structure having
0.113 eV high energy (SO case) than the prism, the face capped pyramid is the 
third isomer with 0.154 eV higher energy, while an out-off plane 
double capped square has been obtained as the last isomer with 0.696 eV relative energy 
to the lowest one. Dissimilar to Pt$_{4}$ and Pt$_{5}$ clusters, SO coupling
effect changes the order of the most Pt$_{6}$ isomers. When SO effect does
not included in the calculations, the second and the third isomers are predicted 
as the face capped pyramid and the bridge capped bipyramid, respectively. Similarly,
3-dimensional double capped tetrahedron has less relative 
energy for non-SO case than SO case. On the other hand, SO effect decreases 
significantly the relative energies of planar or quasi-planar structures. 
For instance, 0.390 eV of septet triangle 
becomes 0.113 eV, 2.031 eV nonet double square becomes 0.334 eV, 0.880 eV of 
double capped square becomes 0.696 eV due to SO mixing. 
The big difference between the relative energies of double square in the nonet
state (2.031 eV for non-SO case and 0.334 eV for SO case) can be explained by
the fact that SO effect highly changes the initial spin multiplicity and 
converges to a spin moment of 0.74 $\mu_{B}$.
Thus, we agree with Huda {\it et al}~\cite{Huda} 
that SO coupling do has a considerable effect on these clusters, and it increases
the stability of planar structures. However, we do not agree with neither 
Huda {\it et al}~\cite{Huda} nor Bhattacharyya and Majumder~\cite{Bhat} that the 
planar structures are the most stable isomers of small Pt$_{n}$ ($n$=4-6) clusters.
This conclusion is supported by not only the results of Futschek {\it et al}~\cite{Futschek}
but also the findings of Tian {\it et al}~\cite{Tian},
who have calculated the structural properties of Pt$_{7}$ cluster by using DFT
with both Gaussian and plane wave basis sets and obtained a 3-dimentional 
coupled tetragonal pyramid as the global minimum, which can be constructed
by adding an atom on the center of one of the rectangular faces of the triangular 
prism. As in the case of Pt$_{5}$, 
while SO effect increases some of the bond lengths of Pt$_{6}$ clusters, it
decreases some others. 

\subsection{Most Stable Isomers of Pt$_{4}$-Pt$_{6}$}

Point group symmetries, spin magnetic moments, total binding energies per atom, 
SO energies per atom
(binding energy difference between non-SO and SO cases), HOMO-LUMO gaps, 
and lowest and highest vibrational frequencies for the most stable spin 
states of each 
isomers of all Pt clusters studied in this work can be found in Table~\ref{table-all}.  
SO coupling effect always reduces binding energy since it makes a larger 
contribution to the atomic energy than to the cluster energy. 
As the cluster size increases from 4 to 6, the binding energy per atom increases
as well (from 2.122 eV to 2.483 eV for non-SO case,
from 1.914 eV to 2.286 eV for SO case). In contrast to the results of
Huda {\it et al}~\cite{Huda}, SO energy per atom decreases from 0.208 eV to
0.193 eV when the size of the clusters changes from 4 to 5. For the lowest energy
structure of Pt$_{6}$, SO energy per atom is 0.196 eV. Except the fisrt isomers 
of Pt$_{5}$ and Pt$_{6}$ and the seventh isomer of Pt$_{6}$, HOMO-LUMO gaps are 
reduced due to SO mixing.  
For the non-SO case, planar or quasi-planar (4-4), (5-5), (5-6), (6-2), (6-4), 
(6-8) structures have significantly large HOMO-LUMO gaps. SO effects reduce
these large gaps. In general, SO coupling effect 
does not change the vibrational frequencies considerably. 

\section{SUMMARY}

In conclusion, we have studied the effect of SO coupling on small Pt clusters,
Pt$_{n}$ ($n$=4-6). Four isomers of Pt$_{4}$, six isomers of Pt$_{5}$, and 
eleven isomers of Pt$_{6}$ were calculated with and without SO effects. 
It is found that SO coupling effects have a considerable effect on these clusters
which can change the order of isomers. Although it increases
the stability of planar structures, it cannot make these planar structures 
the most stable isomers. The lowest energy structures of Pt$_{4}$, 
Pt$_{5}$, and Pt$_{6}$ clusters are predicted as 3-dimentional tetrahedron, pyramid, 
and trigonal prism, respectively. In general, SO mixing reduces both total binding
energies and HOMO-LUMO gaps.

%\begin{acknowledgements}
%\end{acknowledgements}

%\newpage

\newpage

\begingroup
\squeezetable
\begin{table}
\caption{\label{table-pt4}Relative Stability of Pt$_4$ Isomers Predicted by Scalar-Relativistic (non-SO)
and full-relativistic (SO) DFT calculations.}
\begin{ruledtabular}
\begin{tabular}{l c c c c c }
Isomer& Structure   & \multicolumn{2}{c}{non-SO}                & \multicolumn{2}{c}{SO}            \\
                                   \cline{3-4}                    \cline{5-6}
      &             & Spin         & $\Delta$E                       & Spin      & $\Delta$E \\
      &             & Moment ($\mu_{B}$)       & (eV)          & Moment ($\mu_{B}$)   &  (eV)     \\
\hline
(4-1) & Tetrahedron & 2            &    0.000                        & 1.84      & 0.026       \\
      &             & 4            &    0.109\footnote{saddle point} & 3.64      & 0.000       \\
      &             & 6            &    1.110                        & 3.64      & 0.000       \\
      &             & 8            &    2.239                        & 3.64      & 0.000       \\
      &             &              &                                 &           &             \\
(4-2) & Rhombus     & 0            &    0.576                        & 2.08      & 0.133       \\
      &             & 2            &    0.471                        & 2.11      & 0.133       \\
      &             & 4            &    0.206                        & 2.10      & 0.133       \\
      &             & 6            &    0.945                        & 2.09      & 0.134       \\
      &             &              &                                 &           &             \\
(4-3) & Square      & 0            &    0.232                        & 0.00      & 0.357       \\
      &             & 2            &    0.321                        & 1.77      & 0.493       \\
      &             &              &                                 &           &             \\
(4-4) & Y-like      & 0            &    1.410                        & 0.54      & 0.410       \\
      &             & 2            &    0.892                        & 1.50      & 0.458       \\
      &             & 4            &    0.551                        & 3.51      & 0.354       \\
      &             & 6            &    1.134                        & 3.47      & 0.337       \\
\end{tabular}
\end{ruledtabular}
\end{table}
\endgroup

\begingroup
\squeezetable
\begin{table}
\caption{\label{table-pt5}Relative Stability of Pt$_5$ Isomers Predicted by Scalar-Relativistic (non-SO)
and full-relativistic (SO) DFT calculations.}
\begin{ruledtabular}
\begin{tabular}{l c c c c c }
Isomer& Structure   & \multicolumn{2}{c}{non-SO}  & \multicolumn{2}{c}{SO}            \\
                                   \cline{3-4}      \cline{5-6}
      &             & Spin              & $\Delta$E    & Spin              & $\Delta$E \\
      &             & Moment ($\mu_{B}$)& (eV)         & Moment ($\mu_{B}$)&  (eV)     \\
\hline
(5-1) & Pyramid     & 2                 & 0.460        & 3.46              & 0.530\footnote{saddle point}       \\
      &             & 4                 & 0.149        & 3.79              & 0.144       \\
      &             & 6                 & 0.011        & 5.58              & 0.000  \\
      &             & 8                 & 0.816        & 7.28              & 0.757\footnotemark[1]       \\
      &             &                   &              &                   &             \\
(5-2) & Capped Tetrahedron & 0          & 0.920        & 0.00              & 0.855       \\
      &             & 2                 & 0.428        & 2.38              & 0.500       \\
      &             & 4                 & 0.000        & 3.67              & 0.167       \\
      &             & 6                 & 0.336        & 5.18              & 0.278       \\
      &             &                   &              &                   &             \\
(5-3) & W-like      & 0                 & 0.661        & 0.00              & 0.609       \\
      &             & 4                 & 0.169        & 3.58              & 0.173       \\
      &             & 6                 & 0.536        & 2.03              & 0.193       \\
      &             &                   &              &                   &             \\
(5-4) & Bipyramid   & 0                 & 1.011        & 0.02              & 0.807       \\
      &             & 2                 & 0.319        & 2.05              & 0.339\footnotemark[1]       \\
      &             & 4                 & 0.186        & 3.56              & 0.247       \\
      &             & 6                 & 0.609        & 3.57              & 0.247\footnotemark[1]       \\
      &             & 8                 & 0.954        & 3.97              & 0.388       \\
      &             &                   &              &                   &             \\
(5-5) & Capped Square & 2               & 0.674        & 2.43              & 0.363       \\
      &             &                   &              &                   &             \\
(5-6) & X-like      & 0                 & 1.319        & 0.00              & 0.713       \\
      &             & 2                 & 0.930        & 1.83              & 0.524       \\
      &             & 8                 & 2.859        & 3.39              & 0.490       \\
\end{tabular}	
\end{ruledtabular}
\end{table}
\endgroup

\begingroup
\squeezetable
\begin{table}
\caption{\label{table-pt6}Relative Stability of Pt$_6$ Isomers Predicted by Scalar-Relativistic (non-SO)
and full-relativistic (SO) DFT calculations.}
\begin{ruledtabular}
\begin{tabular}{l c c c c c }
Isomer& Structure   & \multicolumn{2}{c}{non-SO}                & \multicolumn{2}{c}{SO}            \\
                                   \cline{3-4}                    \cline{5-6}
      &             & Spin         & $\Delta$E                   & Spin      & $\Delta$E \\
      &             & Moment ($\mu_{B}$)      & (eV)             & Moment ($\mu_{B}$)    &  (eV)     \\
\hline
(6-1) & Trigonal Prism   &      4  & 0.246                       & 3.68      & 0.285       \\
      &             &           6  & 0.000                       & 5.43      & 0.000        \\
      &             &           8  & 0.168                       & 5.41      & 0.001        \\
      &             &              &                             &           &              \\
(6-2) & Triangle    &          2   & 0.923                       & 1.63      & 0.219       \\
      &             &          4   & 0.957\footnote{saddle point}& 2.69      & 0.209       \\
      &             &          6   & 0.390                       & 3.44      & 0.113       \\
      &             &          8   & 1.831                       & 3.43      & 0.115        \\
      &             &              &                             &           &             \\
(6-3) & Face Capped Pyramid &  2   & 0.404                       & 2.50      & 0.333     \\
      &             &          4   & 0.221                       & 4.75      & 0.156     \\
      &             &          6   & 0.150                       & 4.75      & 0.154       \\
      &             &              &                             &           &             \\
(6-4) & Double Square       &  6   & 0.684                       & 5.38      & 0.756       \\
      &                     &  8   & 2.031                       & 0.74      & 0.334       \\
      &             &              &                             &           &             \\
(6-5) & Bridge Capped Pyramid& 0   & 0.865                       & 0.00      & 0.781       \\
      &             &          6   & 0.644                       & 5.42      & 0.338       \\
      &             &          8   & 0.899                       & 5.13      & 0.467       \\
      &             &              &                             &           &             \\
(6-6) & Face Capped Bipyramid & 0  & 1.418                       & 0.00      & 1.044       \\
      &             &           2  & 0.598                       & 1.74\footnote{convergence cannot be achieved}& 0.686\footnotemark[2]       \\
      &             &           8  & 0.369                       & 7.01\footnotemark[2]& 0.355\footnotemark[2]       \\
      &             &              &                             &           &             \\
(6-7) & Octahedron  &          0   & 1.326                       & 0.00      & 1.620       \\
      &             &          2   & 1.132                       & 4.83      & 0.509       \\
      &             &          6   & 1.041                       & 2.44      & 0.711       \\
      &             &              &                             &           &             \\
(6-8) & W-like      &          0   & 0.728                       & 0.00      & 0.763\footnotemark[1]       \\
      &             &          2   & 1.301                       & 4.02      & 0.520       \\
      &             &          6   & 0.530                       & 4.06      & 0.519        \\
      &             &              &                             &           &        \\
(6-9) & Double Capped Tetrahedron&4& 0.398                       & 3.66      & 0.649       \\
      &             &            6 & 0.782                       & 5.08      & 0.678       \\
      &             &              &                             &           &             \\
(6-10)& Bridge Capped Bipyramid &0 & 0.744                       & 0.00      & 0.663       \\
      &             &            4 & 0.349                       & 3.54\footnotemark[2] & 0.801\footnotemark[2]       \\
      &             &              &                             &           &        \\
(6-11)& Double Capped Square & 0   & 0.941                       & 1.31      & 0.807       \\
      &             &          2   & 0.669                       & 1.83      & 0.729       \\
      &             &          4   & 0.728                       & 3.22      & 0.696\footnotemark[1]       \\
      &             &          6   & 0.880                       & 3.22      & 0.696       \\
      &             &          8   & 1.483                       & 3.21      & 0.696       \\
\end{tabular}
\end{ruledtabular}
\end{table}
\endgroup

\begingroup
\squeezetable
\begin{table}
\caption{\label{table-all}Isomeric structure properties of Pt$_n$
($n$=4-6) clusters with and without SO coupling.}
\begin{ruledtabular}
\begin{tabular}{l c c c c c c c c c c c c}
     & Structure      & \multicolumn{2}{c}{Symmetry}  & \multicolumn{2}{c}{Spin}         & \multicolumn{2}{c}{Total BE}  & SO Energy &
     \multicolumn{2}{c}{HOMO-LUMO} & \multicolumn{2}{c}{$\omega_l$ and $\omega_h$\footnote{Lowest and highest vibrational frequencies}}\\
      &             & \multicolumn{2}{c}{}          & \multicolumn{2}{c}{Moment ($\mu_{B}$)}  & \multicolumn{2}{c}{(eV/atom)} & (eV/atom)&  
      \multicolumn{2}{c}{gap (eV)} & \multicolumn{2}{c}{(cm$^{-1}$)} \\
      \cline{3-4} \cline{5-6} \cline{7-8} \cline{10-11} \cline{12-13} 
      &                     &  non-SO   & SO        & non-SO & SO      & non-SO & SO      &        & non-SO\footnote{$\alpha$ spin}& SO    & non-SO  & SO  \\
\hline
(4-1) & Tetrahedron         &  $C_{3v}$ & $D_{2d}$  & 2      & 3.64    & 2.122  & 1.914   & 0.208  & 1.603       & 1.162 & 80, 215 & 99, 215  \\
(4-2) & Rhombus             &  $C_{2v}$ & $C_{2v}$  & 4      & 2.10    & 2.071  & 1.882   & 0.190  & 1.890       & 1.433 & 50, 213 & 44, 206  \\
(4-3) & Square              &  $D_{4h}$ & $D_{4h}$  & 0      & 0.00    & 2.064  & 1.825   & 0.239  & 1.577       & 1.487 & 9, 198  & 41, 187  \\
(4-4) & Y-like              &  $C_{s}$  & $C_{1}$   & 4      & 3.49    & 1.985  & 1.830   & 0.155  & 2.316       & 1.903 & 42, 234 & 48, 241  \\
      &                     &           &           &        &         &        &         &        &             &       &         &   \\
(5-1) & Pyramid             &  $C_{2v}$ & $C_{2v}$  & 6      & 5.58    & 2.319  & 2.126   & 0.193  & 1.584       & 1.629 & 47, 209 & 51, 197  \\
(5-2) & Capped Tetra.       &  $C_{2v}$ & $C_{s}$   & 4      & 3.67    & 2.321  & 2.090   & 0.231  & 1.656       & 1.403 & 39, 212 & 46, 198  \\
(5-3) & W-like              &  $C_{1}$  & $C_{1}$   & 4      & 3.58    & 2.288  & 2.087   & 0.201  & 1.780       & 1.563 & 41, 229 & 46, 230  \\
(5-4) & Bipyramid           &  $D_{3}$  & $C_{1}$   & 4      & 3.56    & 2.284  & 2.077   & 0.207  & 1.807       & 1.236 & 71, 223 & 69, 210  \\
(5-5) & Capped Square       &  $C_{1}$  & $C_{1}$   & 2      & 2.43    & 2.187  & 2.053   & 0.134  & 1.946       & 1.296 & 50, 229 & 50, 228  \\
(5-6) & X-like              &  $C_{1}$  & $C_{1}$   & 2      & 3.39    & 2.135  & 2.028   & 0.107  & 2.063       & 1.754 & 30, 269 & 34, 246  \\
      &                     &           &           &        &         &        &         &        &             &       &         &   \\
(6-1) & Trigonal Prism      &  $C_{2v}$ & $C_{1}$   & 6      & 5.42    & 2.483  & 2.286   & 0.196  & 1.000       & 1.173 & 19, 197 & 11, 184 \\
(6-2) & Triangle            &  $C_{1}$  & $C_{1}$   & 6      & 3.44    & 2.418  & 2.267   & 0.151  & 3.081       & 1.392 & 37, 239 & 26, 242 \\
(6-3) & Face Capped Pyr.    &  $C_{1}$  & $C_{2}$   & 6      & 4.75    & 2.458  & 2.260   & 0.198  & 1.339       & 1.319 & 39, 206 & 38, 201 \\
(6-4) & Double Square       &  $C_{2v}$ & $C_{1}$   & 6      & 0.74    & 2.369  & 2.230   & 0.139  & 1.958       & 1.343 & 31, 230 & 34, 245 \\
(6-5) & Bridge Capped Pyr.  &  $C_{1}$  & $C_{1}$   & 6      & 5.42    & 2.376  & 2.227   & 0.147  & 1.662       & 1.592 & 32, 216 & 22, 204 \\
(6-6) & Face Capped Bipyr.  &  $C_{1}$  & $C_{1}$   & 8      & 7.01    & 2.421  & 2.226   & 0.195  & 2.579       & 1.325 & 35, 200 & \footnote{convergence cannot be achieved} \\
(6-7) & Octahedron          &  $C_{1}$  & $C_{2}$   & 6      & 4.83    & 2.309  & 2.201   & 0.108  & 0.870       & 1.304 & 72, 209 & 49, 194 \\
(6-8) & W-like              &  $C_{1}$  & $C_{s}$   & 6      & 4.04    & 2.395  & 2.199   & 0.196  & 2.444       & 1.384 & 36, 238 & 47, 226 \\
(6-9) & Double Capped Tetra.& $D_{2d}$& $D_{2d}$    & 4      & 3.66    & 2.417  & 2.177   & 0.240  & 1.824       & 1.265 & 28, 192 & 27, 196 \\
(6-10)& Bridge Capped Bipyr.&  $C_{1}$  & $C_{1}$   & 4      & 0.00    & 2.425  & 2.175   & 0.250  & 1.625       & 1.326 & 41, 217 & 48, 217  \\
(6-11)& Double Capped Square&  $C_{2}$  & $C_{1}$   & 2      & 3.22    & 2.371  & 2.170   & 0.202  & 1.511       & 1.279 & 30, 216 & 30, 229 \\
\end{tabular}
\end{ruledtabular}
\end{table}
\endgroup

\end{document}